\documentclass[prl,superscriptaddress,twocolumn]{revtex4}
\usepackage{hyperref}
\usepackage{graphicx,epsfig}
\usepackage{subeqn}
\usepackage[usenames]{color}

\newcommand{\bk}{{\bf k}}

\newcommand{\cP}{{\mathcal P}}

\newcommand{\cZ}{{\mathcal Z}}

\newcommand{\kB}{{k_{\rm B}}}

\begin{document}

\title{From Fractional Exclusion Statistics Back to Bose and Fermi Distributions}
\date{\today}

\author{Drago\c s-Victor Anghel}
\email{dragos@theory.nipne.ro}
\affiliation{Horia Hulubei National Institute for Physics and Nuclear Engineering, P.O. Box MG-6, 077126 M\u agurele, Ilfov, Romania}

\begin{abstract}
Fractional exclusion statistics (FES) is a generalization of the Bose and Fermi statistics. Typically, systems of interacting particles are described as ideal FES systems and the properties of the FES systems are calculated from the properties of the interacting systems. In this paper I reverse the process and I show that a FES system may be described in general as a gas of quasiparticles which obey Bose or Fermi distributions; the energies of the newly defined quasiparticles are calculated starting from the FES equations for the equilibrium particle distribution. In the end I use a system in the effective mass approximation as an example to show how the procedure works.
\end{abstract}

\maketitle

\section{Introduction} \label{intro}

The concept of fractional exclusion statistics (FES) was introduced by Haldane in Ref. \cite{PhysRevLett.67.937.1991.Haldane} and the statistical mechanics of FES systems was formulated by several authors, employing different methods \cite{PhysRevD.45.4706.1992.Ramanathan,PhysRevLett.72.600.1994.Veigy,PhysRevLett.73.922.1994.Wu,PhysRevLett.73.2150.1994.Isakov,PhysRevLett.73.3331.1994.Murthy,FES_intro2013.Murthy,PhysLettB365.202.1996.Polychronakos}.
The FES have been applied to the description of several types of interacting particle systems
\cite{PhysRevLett.67.937.1991.Haldane,PhysRevD.45.4706.1992.Ramanathan,PhysRevLett.72.600.1994.Veigy,PhysRevLett.73.922.1994.Wu,PhysRevLett.73.2150.1994.Isakov,PhysRevLett.73.3331.1994.Murthy,FES_intro2013.Murthy,NewDevIntSys.1995.Bernard,PhysRevB.56.4422.1997.Sutherland,PhysRevE.84.021136.2011.Liu,PhysRevE.85.011144.2012.Liu}
and the properties of FES systems in different numbers of dimensions and trapping potentials have been studied \cite{PhysRevE.75.61120.2007.Potter,PhysRevE.76.61112.2007.Potter,PhysLettA.371.389.2007.Sevincli,PhysRevLett.74.3912.1995.Sen,JPhysB33.3895.2000.Bhaduri,PhysRevLett.86.2930.2001.Hansson,PhysRevE.78.021127.2008.Mirza,PhysRevE.80.011132.2009.Mirza,PhysRevE.82.031137.2010.Mirza,JPhysB.43.055302.2010.Qin,PhysRevE.83.021111.2011.Qin,PhysLettA.376.1191.2012.Qin,CommunTeorPhys.58.573.2012.Qin,PhysRevE.76.061123.2007.Pellegrino,JPhysA.45.315302.2012.vanZyl,JPhysA.46.045001.2013.MacDonald}.
A general method for the description of Fermi liquid type of systems as ideal FES gases have been proposed in Ref. \cite{PhysLettA.372.5745.2008.Anghel,PhysLettA.376.892.2012.Anghel} and have been applied in Refs. \cite{RJP.54.281.2009.Anghel,PhysScr.2012.014079.2012.Anghel,JPhysConfSer.410.012120.2013.Nemnes,arXiv13035493.Anghel}.

Typically, for systems described in the mean field approximation, like the ones analyzed in Refs. \cite{PhysLettA.372.5745.2008.Anghel,PhysLettA.376.892.2012.Anghel,RJP.54.281.2009.Anghel,PhysScr.2012.014079.2012.Anghel,JPhysConfSer.410.012120.2013.Nemnes,arXiv13035493.Anghel}, one employs Landau's Fermi liquid formalism in which quasiparticle energies are defined in such a way that the equilibrium populations are Bose or Fermi distributions (depending on whether we have bosons or fermions in the system) over the quasiparticle energies. The quasiparticle system is neither ideal, nor it is thermodynamically equivalent with the initial interacting particle system, but if one properly redefines the quasiparticle energies, one may transform it into an ideal gas, thermodynamically equivalent with the original system \cite{PhysScr.2012.014079.2012.Anghel}. The ideal gas thus obtained obeys FES. 

The connection between FES and Fermi liquid theory was investigated in more detail in Ref. \cite{arXiv13035493.Anghel} where it was shown that the FES and Landau's quasiparticle populations are identical if one makes the correspondence between the quasiparticle energies in both formalisms.
Here I reverse the process and I describe a FES distribution as a Fermi or Bose distribution over a new set of quasiparticle energies. The quasiparticle energies are determined from the equations for the FES equilibrium particle distribution.

The paper is organized as follows. First I present FES and the equations that give the equilibrium particle populations. In these equations I introduce the ansatz for Bose or Fermi distributions 
and I obtain a set of equations for the quasiparticle energies. I particularize these equations for some common cases. In the end I give the conclusions.

\section{Particle distribution in FES}

For the calculation of the particle distribution in FES one can adopt three equivalent descriptions: bosonic, fermionic \cite{EPL.90.10006.2010.Anghel,JPhysA.40.F1013.2007.Anghel,PhysScr.2012.014079.2012.Anghel} and the standard FES description \cite{PhysRevLett.73.922.1994.Wu}.

\paragraph{The bosonic description.}

We have a general FES system of species $(G_i,N_i,\epsilon_i)$, where $G_i$ is the number of states, $N_i$ is the number of particles, and $\epsilon_i$ is the single-particle energy in the species. The FES parameters, $\alpha^{(-)}_{ij}$, describe the change of the number of states, $\delta G_i$ at a change of the number of particles, $\delta N_j$: $\delta G_i=-\alpha^{(-)}_{ij}\delta N_j$ for any $i$ and $j$. The grandcanonical partition function is
\begin{equation}
  \cZ^{(-)} = \sum_{\{(G_i,N_i)\}}\cZ^{(-)}_{\{(G_i,N_i)\}} \label{cZ_gen}
\end{equation}
and if the particles are bosons the partial partition function, $\cZ^{(-)}_{\{(G_i,N_i)\}}$, is \cite{EPL.90.10006.2010.Anghel,JPhysA.40.F1013.2007.Anghel}
\begin{eqnarray}
  \cZ^{(-)}_{\{(G_i,N_i)\}} &=& \prod_i \frac{(G_i+N_i-1)!}{N_i!(G_i - 1)!} e^{-\beta(\epsilon_i-\mu) N_i} \label{cZB_def}
\end{eqnarray}

If we maximize $\cZ^{(-)}_{\{(G_i,N_i)\}}$ subject to the variation of the species populations we obtain the equations for the equilibrium particle distribution \cite{EPL.90.10006.2010.Anghel,JPhysA.40.F1013.2007.Anghel},
\begin{eqnarray}
  0 &=& \frac{\partial\log\cZ^{(-)}_{\{(G_i,N_i)\}}}{\partial N_k} = \ln\frac{1+n^{(-)}_k}{n^{(-)}_k} - \sum_i\alpha^{(-)}_{ik}\ln(1+n^{(-)}_i) \nonumber \\
  && - \beta(\epsilon_k-\mu) \label{Eq_ni_B}
\end{eqnarray}
where $n_i\equiv N_i/G_i$

\paragraph{The fermionic description.}

Another way to look at the same problem is to assume that while $G_i$ is the number of available states in the species $i$, the actual number of states is $T_i\equiv G_i+N_i$ -- like in the situation when there are $N_i$ fermions on $T_i$ states. In such a case we define the FES parameters, $\alpha^{(+)}_{ij}$, so that $\delta T_i=-\alpha^{(+)}_{ij}\delta N_j$. The partial partition function is
\begin{eqnarray}
  \cZ^{(+)}_{\{(T_i,N_i)\}} &=& \prod_i \frac{T_i!}{N_i!(T_i - N_i)!} e^{-\beta(\epsilon_i-\mu) N_i} \label{cZF_def}
\end{eqnarray}
and the maximization with respect to $N_k$ gives 
\begin{eqnarray}
  0 &=& 
  \ln\frac{1-n^{(+)}_k}{n^{(+)}_k} + \sum_i\alpha_{ik}\ln(1-n^{(+)}_i) 
  - \beta(\epsilon_k-\mu) , \label{Eq_ni_F}
\end{eqnarray}
where $n^{(+)}_i=N_i/T_i$. 

The fermionic description is more appropriate for FES in Fermi systems and changes into the bosonic description by the redefinitions $\alpha^{(-)}_{ij}\equiv \delta_{ij}+\alpha^{(+)}_{ij}$, $G_i\equiv T_i-N_i$ and $n^{(+)}_i =n^{(-)}_i /(1-n^{(-)}_i)$.

\paragraph{The general FES description.}

The usual description of FES systems was originally proposed by Wu \cite{PhysRevLett.73.922.1994.Wu}. In this case one defines the number of states in the species $i$ when no particles are in the system ($N_i=0$ for any $i$): $G^{0}_i\equiv G_i + \sum_j\alpha^{(-)}_{ij} N_j$. Writing the partial partition function (\ref{cZB_def}) in terms of $G^{0}_i$ and defining $n^{(0)}_i\equiv N_i/G^{(0)}_i = n^{(-)}_i/[1+\sum_j (\alpha^{(-)}_{ij} n^{(-)}_j G_j/G_i) ]$, by the maximization procedure one obtains the system of equations,
\begin{subequations}\label{syst_Wu}
\begin{eqnarray}
    (1+w_{i}) \prod_{j}\left( \frac{w_{j}}{1+w_{j}} \right)^{\alpha^{(-)}_{ji}}
    &=& e^{(\epsilon_{i}-\mu)/\kB T}, \label{NLS} \\
  \sum_j(\delta_{ij}w_j + \alpha_{ij}G^{(0)}_{j}/G^{(0)}_{i})n^{(0)}_j &=& 1. \label{LS}
\end{eqnarray}
\end{subequations}
Equations (\ref{NLS}) should be solved to obtain Wu's auxiliary functions, $w_i$, which then may be plugged into Eqs. (\ref{LS}) to calculate the equilibrium populations, $n^{(0)}_j$. Notice that $w_i\equiv 1/n^{(-)}_i$ \cite{JPhysA.40.F1013.2007.Anghel}. 

\paragraph{The FES distribution written as a Bose or a Fermi distribution.}

We look for solutions of the form 
\begin{equation}
  n^{(\pm)}_i = \frac{1}{e^{\beta(\tilde\epsilon_i-\mu)}\pm 1} , \label{ansatz_ni}
\end{equation}
for the Eqs. (\ref{Eq_ni_B}) and (\ref{Eq_ni_F}) -- in what follows we shall always use the upper signs for fermions and the lower signs for bosons. Plugging (\ref{ansatz_ni}) into Eqs. (\ref{Eq_ni_B}) and (\ref{Eq_ni_F}) we obtain a self-consistent set of equations for the quasiparticle energies,
\begin{eqnarray}
  \tilde\epsilon_k &=& \epsilon_k \mp k_BT \sum_i\alpha_{ik}\ln\left[1\mp n^{(\pm)}_i\right] . \nonumber \\
  &=& \epsilon_k \pm k_BT \sum_i\alpha_{ik}\ln\left[1\pm e^{-\beta(\tilde\epsilon_i-\mu)} \right] \label{eq_tilmu1}
\end{eqnarray}
%

\begin{figure}[t]
  \centering
  \includegraphics[width=8cm,keepaspectratio=true]{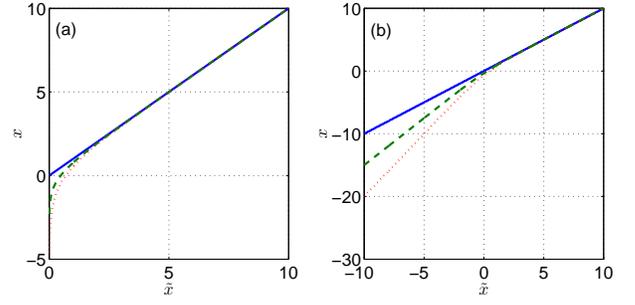}
  \caption{(Color online) The variation of $x$ with respect to $\tilde x$ for $\alpha_{ij}=\alpha\delta_{ij}$ and $\alpha = 0,0.5,1$, from up to down (solid-blue, dashed-green and dotted-red lines, respectively) for bosons (a) and fermions (b).}
  \label{FES-FLT-x_vs_tilx}
\end{figure}

In many situations, $\alpha_{ij} \equiv \alpha\delta_{ij}$ (e.g. \cite{PhysRevLett.73.3331.1994.Murthy,PhysRevLett.74.3912.1995.Sen,JPhysB33.3895.2000.Bhaduri,PhysRevLett.86.2930.2001.Hansson,JPA35.7255.2002.Anghel,RJP.54.281.2009.Anghel}). In such a case
\begin{subequations}\label{eq_tilmu1_adiagT}
\begin{eqnarray}
  \tilde\epsilon_k &=& \epsilon_k \pm \alpha k_BT \ln\left[1\pm e^{-\beta(\tilde\epsilon_k-\mu)} \right] , \label{eq_tilmu1_adiag} \\
  &=& \frac{\epsilon_k}{1\pm\alpha} \pm \frac{\alpha}{1\pm\alpha} [\mu - \ln n^{(\pm)}(\epsilon_k)] . \label{eq_tilmu1_adiag2}
\end{eqnarray}
or
\begin{equation}
  x = \tilde x \mp \alpha \ln\left(1 \pm e^{-\tilde x}\right), \label{eq_tilmu1_adiag3}
\end{equation}
\end{subequations}
where $x\equiv\beta(\epsilon-\mu)$ and $\tilde x\equiv\beta(\tilde\epsilon-\mu)$. The dependence of $x$ on $\tilde x$ is plotted in Fig. \ref{FES-FLT-x_vs_tilx}.

In Fig. \ref{FES-FLT-x_vs_tilx}(a) we observe that $\tilde x>0$ -- as it should in order to avoid divergent population of the ground state. Moreover, at high energies $x$ and $\tilde x$ become indistinguishable for both, bosonic and fermionic systems.


Another important situation is the quasi-continuous case with the ansatz \cite{EPL.90.10006.2010.Anghel}
\begin{equation}
\alpha_{ij}=\alpha^{(e)}_{ij}+\alpha^{(s)}_{i}\delta_{ij}, \label{alpha_nd}
\end{equation}
where $\alpha^{(e)}_{ij}\equiv a_{ij}G_i$ are the ``extensive'' parameters. We define the species by dividing the single-particle energy axis, $\epsilon$, into small intervals, $\delta\epsilon_i$, centered on $\epsilon_i$, where $i=0,1,\ldots$. Each such interval represents a species with $G_i=\sigma(\epsilon_i)\delta\epsilon_i$ and $N_i=G_in_i$. Using the ansatz (\ref{alpha_nd}) we obtain the equations \cite{EPL.90.10006.2010.Anghel}
\begin{eqnarray}
  && \beta(\mu-\epsilon)+\ln\frac{[1\mp n^{(\pm)}(\epsilon)]^{1\pm\alpha^{(s)}_{\epsilon}}}{n^{(\pm)}(\epsilon)} = \mp \int_{\epsilon_0}^\infty \sigma(\epsilon') \nonumber \\
  && \times\ln[1\mp n^{(\pm)}(\epsilon')]a_{\epsilon'\epsilon}\,d\epsilon' , \label{inteq_for_n2c}
\end{eqnarray}
where $\sigma(\epsilon)$ is the DOS along the $\epsilon$ axis and $\epsilon_0$ is the lowest particle energy. From Eqs. (\ref{inteq_for_n2c}) and (\ref{ansatz_ni}) we get
\begin{eqnarray}
  \beta(\tilde\epsilon-\epsilon) &=& \mp\alpha^{(s)}_{\epsilon}\ln[1\mp n^{(\pm)}(\epsilon)] \nonumber \\
  && \mp \int_{\epsilon_0}^\infty \sigma(\epsilon')\ln[1\mp n^{(\pm)}(\epsilon')] a_{\epsilon'\epsilon} \,d\epsilon' \nonumber \\
  &=& \pm\alpha^{(s)}_{\epsilon}\ln[1\pm e^{-\beta(\tilde\epsilon-\mu)}] \pm \int_{\tilde\epsilon_0}^\infty \tilde\sigma(\tilde\epsilon') \nonumber \\
  && \times \ln[1\pm e^{-\beta(\tilde\epsilon'-\mu)}] a_{\epsilon(\tilde\epsilon') \epsilon(\tilde\epsilon)} \,d\tilde\epsilon' , \label{inteq_for_qpen}
\end{eqnarray}
where $\tilde\sigma(\tilde\epsilon)$ is the DOS along the $\tilde\epsilon$ axis,
\begin{equation}
  \tilde\sigma(\tilde\epsilon) = \sigma(\epsilon)\left|d\tilde\epsilon/d\epsilon\right|^{-1} , \label{def_til_eps}
\end{equation}
and $\tilde\epsilon_0\equiv\tilde\epsilon(\epsilon_0)$ is the lowest quasiparticle energy.
Equation (\ref{inteq_for_qpen}) becomes identical with Eq. (\ref{eq_tilmu1_adiag3}) if we take $a_{\epsilon' \epsilon} = 0$ for any $\epsilon'$ and $\epsilon$. 

If we define a function $\cP(\tilde\epsilon',\tilde\epsilon)$ such that
\begin{equation}
  \frac{\partial \cP(\tilde\epsilon',\tilde\epsilon)}{\partial\tilde\epsilon'} = \tilde\sigma(\tilde\epsilon') a_{\epsilon(\tilde\epsilon') \epsilon(\tilde\epsilon)} , \label{def_P}
\end{equation}
then Eq. (\ref{inteq_for_qpen}) becomes
\begin{eqnarray}
  \beta(\tilde\epsilon-\epsilon) &=& \pm\alpha^{(s)}_{\epsilon}\ln[1\pm e^{-\beta(\tilde\epsilon-\mu)}] \mp \ln[1\pm e^{-\beta(\tilde\epsilon_0-\mu)}] \nonumber \\ 
  && \times \cP(\tilde\epsilon_0,\tilde\epsilon) + \beta\int_{\tilde\epsilon_0}^\infty \frac{\cP(\tilde\epsilon',\tilde\epsilon)}{e^{\beta(\tilde\epsilon'-\mu)} \pm 1} \,d\tilde\epsilon' \nonumber \\
  &=& \pm\alpha^{(s)}_{\epsilon}\ln[1\pm e^{-\beta(\tilde\epsilon-\mu)}] \mp \ln[1\pm e^{-\beta(\tilde\epsilon_0-\mu)}] \nonumber \\
  && \times \cP(\tilde\epsilon_0,\tilde\epsilon) + \beta\int_{\tilde\epsilon_0}^\infty \frac{\cP(\tilde\epsilon',\tilde\epsilon)}{\tilde\sigma(\tilde\epsilon')} n(\tilde\epsilon') \tilde\sigma(\tilde\epsilon') \,d\tilde\epsilon' . \label{inteq_for_qpen3}
\end{eqnarray}
Equation (\ref{inteq_for_qpen3}) has the typical form of the Landau's quasiparticle energy,
\begin{equation}
  \tilde\epsilon^L(t) = t + \int_0^\infty V_{tt'} n^{(\pm)}(t') \, dt' , \label{def_epsL_c}
\end{equation}
where $t$ is the energy of the free particle (boson or fermion) and $V_{tt'}$ is the interaction energy between two particles in the single-particle states $t$ and $t'$, respectively.

For fermionic systems in the low temperature limit Eq. (\ref{inteq_for_qpen}) may be simplified to
\begin{eqnarray}
  \tilde\epsilon &=& \epsilon - k_BT\alpha_\epsilon^{(s)} \ln[1-n^{(+)}(\epsilon)] \nonumber \\ 
  && +\frac{\pi^2}{12}(k_BT)^2\tilde\sigma(\mu)(a_{\mu^+ \mu} + a_{\mu^-\mu}) \nonumber \\ 
  && - \int_{\epsilon_0}^\mu\sigma(\epsilon')a_{\epsilon'\epsilon}[\tilde\epsilon(\epsilon')-\mu] d\epsilon \label{til_eps_lowT1}
\end{eqnarray}
where $a_{\mu^+ \mu} \equiv\lim_{\epsilon'\searrow\mu}a_{\epsilon' \mu}$, $a_{\mu^- \mu} \equiv \lim_{\epsilon'\nearrow\mu}a_{\epsilon' \mu}$, and $\tilde\sigma(\mu)\equiv\tilde\sigma(\tilde\epsilon=\mu)$. If we neglect the terms proportional to $k_BT$ and $(k_BT)^2$, then
\begin{eqnarray}
  \tilde\epsilon &=& \epsilon - \int_{\epsilon_0}^\mu\sigma(\epsilon')a_{\epsilon'\epsilon}[\tilde\epsilon(\epsilon')-\mu] d\epsilon \label{til_eps_lowT2}
\end{eqnarray}

As an application for this case let's consider a mean-field model with the effective mass approximation. The total energy of the system is
\begin{subequations} \label{defs_Etot_m*}
\begin{equation}
  E = \sum_\bk\frac{\hbar^2 k^2}{2m^*}n_\bk  \label{Etot_def} 
\end{equation}
and
\begin{equation}
  \frac{1}{2m^*} = \frac{1}{2m} - r\frac{N}{V} \equiv \frac{1}{2m} - r\rho , \label{1_m*_def}
\end{equation}
\end{subequations}
where $m^*$ is the effective mass, $N$ is the total particle number, $V$ is the volume, and $\rho$ is the particle density -- $\rho$ takes values only in a range such that $m^*>0$. For simplicity, we assume that the particles have no spin. 

This is a clear manifestation of FES. 
The kinetic energy is $\epsilon\equiv\hbar^2 k^2/(2m^*)$ and in a $d$-dimensional system the DOS along the $\epsilon$ axis is
\begin{equation} \label{sigma_d}
\sigma_d(\epsilon,m^*) = V_d C_d\left(\frac{2m^*}{\hbar^2}\right)^{d/2} \epsilon^{(d/2)-1} ,
\end{equation}
where $C_d \equiv d\left[2^{d+1}\pi^{d/2} \Gamma(d/2+1)\right]^{-1}$. 
At the addition of a particle into the system the whole spectrum changes. This changes the DOS. If the $\epsilon$ axis is divided into the intervals $\delta\epsilon_i$, then the number of states in each of these intervals changes with $N$ and this implies FES. The FES parameters are $\alpha_{\epsilon,\epsilon_i} = -\delta\epsilon [\partial\sigma(\epsilon,N)/\partial N]$, where 
\begin{eqnarray}
  && \frac{\partial\sigma_d(\epsilon,m^*)}{\partial m^*} = \frac{dV}{2 m^*}C_d \left(\frac{2m^*}{\hbar^2}\right)^{d/2}\epsilon^{d/2-1} , \label{dsigma_dmstar} \\
  && \frac{\partial\sigma}{\partial N} = \frac{\partial\sigma}{\partial m^*} \frac{\partial m^*}{\partial N} = \frac{rdm^*}{V} \sigma_d(\epsilon,m^*) . \label{dsigma_sN} 
\end{eqnarray}
Therefore
\begin{equation}
  \alpha_{\epsilon,\epsilon_i} = -\delta\epsilon (rdm^*/V) \sigma_d(\epsilon,m^*) , \label{alpha}
\end{equation}
so $a_{\epsilon,\epsilon_i} = -rdm^*/V$ is a constant and $\alpha_\epsilon^{(s)}=0$.  If we plug these into (\ref{inteq_for_qpen}), we get
\begin{eqnarray}
  \tilde\epsilon &=& \epsilon \pm k_BTC_d\left(\frac{2m^*}{\hbar^2}\right)^{d/2} rdm^* \int_{0}^\infty {\epsilon'}^{(d/2)-1} \nonumber \\ 
  && \times\ln\left\{1\mp e^{-\beta[\tilde\epsilon(\epsilon')-\mu]}\right\} \,d\epsilon' \nonumber \\
  &=& \epsilon + C_d\left(\frac{2m^*}{\hbar^2}\right)^{d/2} 2rm^* \int_{0}^\infty {\epsilon'}^{d/2} n^{(\pm)}(\epsilon')\,d\epsilon' \nonumber \\
  &=& \epsilon + 2rm^*E/V \label{inteq_for_qpen_ex}
\end{eqnarray}
where we noticed that $\partial\tilde\epsilon/\partial\epsilon = 1$ and the function $\cP$ is
\begin{equation}
  \cP(\tilde\epsilon',\tilde\epsilon) = C_d\left(\frac{2m^*}{\hbar^2}\right)^{d/2} 2rm^* \left[\tilde\epsilon' - 2rm^*E/V \right]^{d/2}
\end{equation}

The standard method for the calculation of the thermodynamic properties of the system is (following Landau's procedure) to define the population as
\begin{subequations} \label{sol_FLT}
\begin{equation}
  n^{(\pm)} = \left[1\pm e^{\beta(\tilde\epsilon^{L}-\mu)}\right]^{-1} , \label{pop_Landau}
\end{equation}
where 
\begin{equation}
  \tilde\epsilon^L_k \equiv \frac{\partial E}{\partial n_k} = \epsilon_k +  2rm^*U/V \label{til_epsL}
\end{equation}
\end{subequations}
is Landau's quasiparticle energy and is obtained from Eq. (\ref{defs_Etot_m*}). Now we see that Eqs. (\ref{inteq_for_qpen_ex}) and (\ref{til_epsL}) are identical. 

\section{Conclusions}

In this paper I showed that a fractional exclusion statistics (FES) system may be described as a system of quasiparticles which obey Bose or Fermi statistics. Using the FES equations for the equilibrium particle distribution, I derive the equations for the energies of the newly defined quasiparticles. This is the reverse of the process of describing interacting bosons and fermions as ideal FES particles \cite{PhysLettA.372.5745.2008.Anghel,PhysLettA.376.892.2012.Anghel,arXiv13035493.Anghel} and emphasizes the general analogy and transformation methods between Bose, Fermi, and fractional exclusion statistics.

In the end I used as example a gas of particles in the effective mass approximation and I showed that by the method proposed here I recover the standard Fermi liquid theory solution. 

\section{Acknowledgements}

Discussions with Francesca Gulminelli, Alexandru Nemnes, and Ionel \c Tifrea are gratefully acknowledged.
%
The work was supported by the Romanian National Authority for Scientific Research CNCS-UEFISCDI projects PN-II-ID-PCE-2011-3-0960 and PN09370102/2009. The travel support from the Romania-JINR Dubna collaboration project Titeica-Markov are gratefully acknowledged.



\end{document}